\documentstyle[twoside,fleqn,espcrc2,psfig]{article}

\newcommand{\AmS}{{\protect\the\textfont2
  A\kern-.1667em\lower.5ex\hbox{M}\kern-.125emS}}

\title{The optical light curve of GRB 970228 refined}

\author{N. Masetti, C. Bartolini, A. Guarnieri and A. Piccioni\address{
Dipartimento di Astronomia, Universit\`a di Bologna\\
via Zamboni, 33 I--40126 Bologna, Italy}}
       
\begin{document}

\begin{abstract}
We present the $R$ and $V$
light curves of the optical counterpart of GRB 970228. 
A critical analysis of all the available data is made in light 
of the results achieved in the recent GRB Symposium held in Huntsville and by
considering the latest information from the HST images on the underlying 
nebulosity.
\end{abstract}

\maketitle

\section{INTRODUCTION}

The recent results obtained by the HST on the optical counterpart of GRB 970228
[3,4] have given answer to important questions concerning
its light decay, its proper motion and the nature of the underlying nebula. 
This information must be taken into account in reanalysing all the previously 
published data.  

In this paper we aim to refine the values and lower
limits (particularly those obtained with the 1.5--meter telescope of the
Bologna Astronomical Observatory) of the points in the $R$ light curve of GRB 
970228, taking into account 
the refined magnitude value for the underlying nebulosity given by 
Fruchter et al. [4]. Moreover, we show for the first time and briefly 
discuss the $V$ light curve of this GRB.

\section{THE LIGHT CURVES}

Galama et al. [5] collected all the known $R$ points and 
lower limits referring to the GRB 970228 optical light curve. 
They however used a value for the luminosity of the nebulosity associated to 
the object ($R=24.0$; Groot et al. [6]) which is overestimated 
according to the most recent HST observations ($R=25.3\pm0.3$; 
Fruchter et al. [4]).
This discrepancy is likely due to calibration errors which depend on
the faintness of the object and not on an intrinsic variability; Fox et al. 
[2] and Fruchter et al. [3] actually
showed that the nebula is constant in brightness.
Therefore, we refined the $R$ data collection by Galama et al. [5] 
using the new value of the magnitude for the underlying fuzzy object; 
in particular, since
the CCD observations obtained at Bologna Observatory [7]
measured the total flux received from the 
transient plus the nebula and a nearby 
K--type star whose magnitude is $R=22.4$ [9], we obtained the
new $R$ magnitudes or lower limits for the optical transient by subtracting
the two latter contributions from the data.

We want to note that in the Galama et al.'s [5] data collection the 
$R$ point by Guarnieri et al. [7] is wrongly plotted
and must be corrected. Indeed, the quoted $R$ magnitude refers to
the total contribution coming from the transient plus the nebulosity and the 
nearby star, and not from the optical transient alone.
The right value on Feb. 28.827 (and not on Feb. 28.76) is therefore 
$R=21.5\pm0.3$.

In this revision Pedichini et al.'s [10] observation of Feb. 28.81 was 
not considered because of the difficulty of reliably converting their color 
system to the $R$ band. Anyway,
the importance of Pedichini et al.'s [10] observations lies in the 
fact of providing a variation $\Delta$m $>2.7$ between Feb. 28 and Mar. 
4. Their observations span from
Feb 28.795 to 28.827 UT. The Bologna $R$ frame was taken between Feb. 28.816
and 28.837 UT, so the two observations partially overlapped. The 
magnitude could be quite different only if the luminosity of the transient 
during the first part of the observation was higher than in the last part.
Assuming that Pedichini et al.'s [10] magnitude on Feb. 28 was the same 
as that found in Bologna in the $R$ band the same day, 
the lower limit $R>24.2$ could be derived for the day Mar. 3.8. 
We consider this figure as representative of the magnitude variation in the
$R$ band, too.

From the observation of Margon et al. [8] on Mar. 3.1, made with 
the Astrophysical Research Consortium 3.5--meter telescope at Apache Point
in the APM $b_{\rm J}$ photometric band, Galama et al. [5] 
derived an estimate of $R$ by interpolating between the color
index $B-R$ measured on Feb. 28 and on Mar. 9. 
Bartolini et al. [1] found evidence
of fast color variations near the maxima of the optical counterpart of both
GRB 970228 and GRB 970508, so it could be dangerous to extrapolate the $R$ 
magnitude from Margon et al.'s [8] observation and to assume 
$b_{\rm J}$ = $B$; moreover, Galama et al. [5] used the color index
of the optical transient alone on Feb. 28, and that of the optical transient
plus the underlying extended source on Mar. 9. For these reasons we did not 
include the value obtained by Galama et al. [5] from the Apache Point
measurement.

The refined $R$ light curve of the optical transient associated to GRB 970228 
is shown in Fig. 1.

We also collected from the literature all the $V$ data points
[3,7,11,12] and plotted them in Fig. 2. 
The $V$ band lower limits published by van Paradijs et al. [12] were
slightly corrected with the use of the $V$ magnitude of the nebula 
($25.7\pm0.15$, as reported by Fruchter et al. [3]).
It should be noted that the value by Guarnieri et al. [7] is 
interpolated from their $B$ and $R$ data acquired on Feb. 28.

\begin{figure}[htb]
\psfig{file=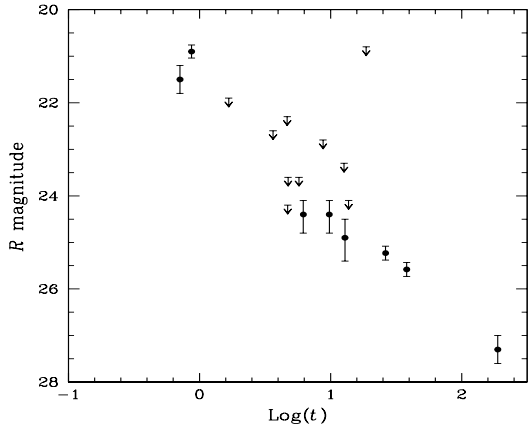,height=6.5cm,width=8cm,angle=0}
\caption{Refined $R$ light curve of the optical transient associated to 
GRB 970228; $t$ is the time (in days) elapsed from the $\gamma$ burst.}
\end{figure}

\begin{figure}[htb]
\psfig{file=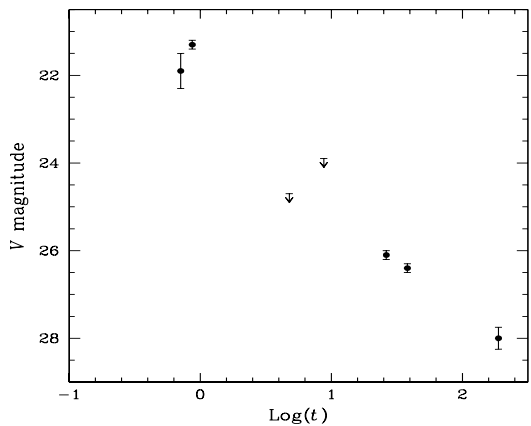,height=6.5cm,width=8cm,angle=0}
\caption{$V$ light curve of the optical transient associated to 
GRB 970228; $t$ is defined as in Fig. 1.}
\end{figure}

\section{DISCUSSION}

The overall trends of the $R$ and $V$ light curves of the transient associated
to GRB 970228 are quite similar. Indeed, we find that both decay following
a power law with spectral indices $\alpha_V=1.27\pm0.08$ and 
$\alpha_R=1.21\pm0.02$, i.e. they are coincident within the errors.
This similarity is however related to the long--term trend. Actually, in a
short time scale the $R$ light curve (Fig. 1) deviates from a unique
power law. The decay from the light peak can indeed be
divided into two phases: the first one, spanning from the optical maximum and 
lasting 3--4 days with $\alpha_R>2.1$ and the second one, after March 4, with 
$\alpha_R<0.6$.
These figures are in agreement with the findings by Galama et al. [5] 
and confirm the rapid light decay noticed by Guarnieri et al. [7]
during the first 3 days after the optical maximum.

We cannot be sure of a similar trend in the $V$ since
at that time the coverage of the light curve was very poor (only two lower
limits are available; see Fig. 2); we see however that the lower limit
of Mar. 4.9 [12] seems to suggest a possible rapid
decay also in $V$.

\bigskip
With the refined value of the $R$ magnitude of the underlying object,
we can now correct the value for the ratio between the luminosity of
the optical transient and that of the fuzzy object in the $R$ band reported 
by Guarnieri et al. [7]. We find that in $R$ the transient at
maximum light was $\approx$60 times brighter than the underlying nebulosity.

If the nebula is a host galaxy, the optical transient associated to 
GRB 970228 has been by far the brightest variable object known up to now.

\section{CONCLUSIONS}

We refined the $R$ light curve of the optical transient associated to GRB
970228 with the use of a more correct value for the magnitude of its 
underlying nebula and compared this curve to the one in the $V$ band 
(previously unpublished).
It is interesting that both the overall trends follow a power law decay 
and are almost identical within the errors. 

The ratio between the $R$ luminosity of the optical transient at maximum 
and the nebulosity has been corrected: now we know that the transient was
approximately 60 times brighter than the nebula at that time.

\bigskip
\noindent
{\it Acknowledgements}. This Investigation was supported by the University of 
Bologna (Funds for selected topics).
We thank D. Lamb and G. Valentini for useful discussions.

\end{document}